\documentclass[pra,showpacs,reprint,longbibliography]{revtex4-1}

\usepackage{epsfig,amsmath,amssymb}
\usepackage[outdir=./]{epstopdf}

\begin{document}
	
\title{Which-way double slit experiments and Born rule violation} 

\author{James Q. Quach} 
\email{quach.james@gmail.com}
\affiliation{ICFO - Institut de Ci\`{e}ncies Fot\`{o}niques, The Barcelona Institute of Science and Technology, 08860 Castelldefels, Spain}

\begin{abstract}
	In which-way double-slit experiments with perfect detectors, it is assumed that having a second detector at the slits is redundant, as it will not change the interference pattern. We however show that if higher-order or non-classical paths are accounted for, the presence of the second detector will have an effect on the interference pattern. Accounting for these non-classical paths also means that the Sorkin parameter in triple-slit experiments is only an approximate measure of Born rule violation. Using the difference between single and double which-way detectors, we give an alternative parameter which is an exact measure of Born rule violation.
\end{abstract}


\maketitle

\section{Introduction}
\begin{figure} [b]
	\centering
		\includegraphics[width=1\columnwidth]{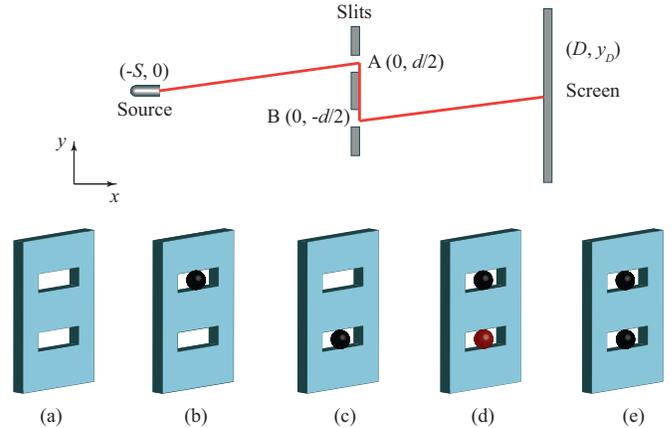}
	\caption{(color online) Top: Side view of one of the myriad of non-classical paths that enters both slits: the path enters slit $A$, makes an abrupt turn, briefly enters slit $B$, before hitting the detecting screen. Bottom: \textit{gedanken} experiments with balls serving as detectors to illustrate different which-way detector setups. (a) setup 1: no which-way detectors, (b) setup 2: a type I which-way detector at slit \textit{A}, (c) setup 3: a type I which-way detector at slit \textit{B}, (d) setup 4: type I which-way detectors at slit \textit{A} and \textit{B}, (e) setup 5: a type II which-way detector.}
	\label{fig:double_slit}
\end{figure}

The which-way double-slit experiment is used as the archetypal experiment to dramatically demonstrate the quantum weirdness of wave-particle duality and wave function collapse: the explanation is that by merely knowing which slit the particle went through collapses the wave function, destroying the wave-like interference effects between the two slits. In the ideal case of perfect detectors, it is generally assumed that whether which-way detectors are placed at one slit or both will yield the same results. The reason for this is that a detection at one slit means that the particle is assumed to not have gone through the other slit. Now on the other hand, in the Feynman path integral formulation of quantum mechanics all possible paths between points contribute to the wave function; this even includes paths that go through one slit then the other as depicted in Fig.~\ref{fig:double_slit}. The inclusion of these \textit{non-classical} or \textit{high-order paths} provides corrections to the interference patterns.

The Born rule is a fundamental axiom of quantum mechanics. It states that if a quantum object is represented by a wave function $\psi(\mathbf{r},t)$, then the probability density of detecting it at position $\mathbf{r}$ and time $t$ is given by the absolute square of the wave function~\cite{born26}, 
\begin{equation}
	P(\mathbf{r},t) = \psi^*(\mathbf{r},t)\psi(\mathbf{r},t) = |\psi(\mathbf{r},t)|^2~.
\label{eq:P}
\end{equation}

Despite being a cornerstone of quantum mechanics, a direct test of the Born rule was not attempted until 2010 by Sinha \textit{et al.}~\cite{sinha10}. The test was a measure of the Sorkin parameter~\cite{sorkin94}, which quantifies non-pairwise interference, in a triple-slit experiment. As the exponent of the Born rule only allows for pairwise interference, a non-zero Sorkin parameter would suggest violation of the Born rule. If beyond-pairwise inference were indeed ever detected, it would likely lead to a modification of the Schr\"{o}dinger's equation, and may importantly provide a sign-post for beyond standard model theories~\cite{barnum14,lee15}. The Sinha \textit{et al.} experiment however found the Sorkin parameter to be zero, within experimental error bounds, and concluded no Born rule violation. Subsequent more precise measurements of the Sorkin parameter also came to the same conclusion~\cite{sollner12,park12,kauten16}. Shortly after the Sinha \textit{et al.} experiment however, it was pointed out that underlying the Sorkin parameter was the assumption that the wave function in the multi-slit setup is simply the superposition of the individual wave functions of the constituent single-slit setups~\cite{raedt12,sawant14}. Strictly speaking this is an approximation, as first pointed out in Ref.~\cite{yabuki86}. Correcting this approximation by including non-classical paths renders the Sorkin parameter non-zero, without violating the Born rule; in fact in some regimes these corrections can be significant~\cite{raedt12,sawant14,sinha15}. In a recent landmark experiment, a non-zero Sorkin parameter due to non-classical paths was indeed measured for the first time in the microwave regime~\cite{rengaraj16}. The Sorkin parameter therefore is not an exact test of Born rule violation.  Given the success of quantum mechanics, any violation of the Born rule is expected to be small. Therefore for a parameter to be a useful measure of Born rule violation, it is important that it is accurate to higher-orders. One notes that alternatives to slit experiments, such as the single spin experiment~\cite{jin17}, may test the Born rule without the
reliance on spatial interference.

Here we will show that by accounting for non-classical paths, the which-way double-slit experiment with one and two which-way detectors will produce different inference patterns, contrary to commonly held assumptions. Making use of this difference, we give an alternative parameter that completely accounts for higher-order corrections, to exactly test the Born rule: the parameter will be exactly zero if the Born rule is not violated, non-zero otherwise. We first will consider the case of perfect detectors and then generalise to imperfect detectors.

\section{Perfect Detectors}
\label{sec:perfect}

\subsection{Which-way double-slit experiment}

Let us consider a double-slit experiment with two types of which-way detectors: one detects whether a particle has gone through slit $A$ or $B$, the other detects that a particle has gone through one or both slits, but does not know which one. We will assume perfect detectors (Sec.~\ref{sec:imperfect} generalises to imperfect detectors). Such detectors can be illustrated in a \textit{gedanken} experiment, with light balls placed precariously in the slits to serve as the detectors (Fig.~\ref{fig:double_slit}). In one type of detector a ball is placed in slit \textit{A}, such that a particle entering the slit will cause it to fall into a tray. We look into the tray to reveal which slit the particle went through: if we see a ball in the tray then the particle went through slit \textit{A}, if we do not see a ball, then the particle went through slit \textit{B}. Implicit here is the assumption that we can not directly view the ball until it hits the tray; in this sense the tray can be thought of as representing a signal amplifier. In another detector, indistinguishable-balls are placed in each slit. In this case, peering into the tray we will see either one or two balls. One ball indicates that the particle had gone through one slit but it does not reveal which one, whereas two balls indicate that the particle had gone through both slits. We will call the former type I and latter type II which-way detectors. Such types of detectors have  been realised in neutron~\cite{summhammer87} and molecular~\cite{cronin09} interference setups, electrons in semiconductors~\cite{aleiner97}, atomic double-pulse Ramsey interferometer experiments~\cite{bertet01}, inelastic electron holography~\cite{potapov06}, electron interferometers~\cite{sonnentag07}, and with ion and electron beam nanofabrication~\cite{frabboni10}.

If $\psi_A$ represents the wave function from a single slit $A$ and $\psi_B$ from a single slit $B$, it is widely taught that the intensity or probability distribution in the double slit experiment is $P_{AB} = |\psi_A + \psi_B|^2$. The correction to this approximation can be quantified with the Feynman path integral formulation. In this formulation all paths between points are possible, even paths that are vastly different from classical paths (classical paths extremise the classical action). One of the myriad of non-classical paths that enter both slits is depicted in Fig.~\ref{fig:double_slit}. As these type of paths enter both slits, they are not captured by the individual single-slit wave functions, $\psi_A$ and $\psi_B$. If we label the contribution from paths that go through both slit with $\psi_{AB}$, the probability distribution for the double-slit experiment with no which-path detectors [Fig.~\ref{fig:double_slit}(a): setup 1] is corrected to
\begin{equation}
	P_{AB} = |\psi_A + \psi_B + \psi_{AB}|^2~.
	\label{eq:PAB}
\end{equation}
The higher-order corrections are typically small, but can be significant~\cite{raedt12,sawant14,sinha15,rengaraj16}. These corrections are not exclusive to the quantum mechanics, but are also present using Maxwell's equations as numerically calculated by Ref.~\cite{raedt12}. Note that $\psi_{A(B)}$ represents all paths that go through slit $A(B)$ only, including paths that go through that slit multiple times, and $\psi_{AB}$ represents all paths that go through both slits including paths that enter the slits multiple times.

Now let us place a type I which-path detector at slit A [Fig.~\ref{fig:double_slit}(b): setup 2]. Conventionally, a detection at slit $A$ means that the particle did not go through slit $B$, otherwise the particle went through slit $B$, so that the probability density is $P_{D_A} =|\psi_A|^2+|\psi_B |^2.$  This however does not account for non-classical paths which can go through slit $A$ and $B$. If one accounts for non-classical paths, than a detection at slit $A$ includes paths that only go through slit $A$ as well as paths that go through slit $A$ and $B$; as they are indistinguishable, we must sum both types of paths ($\psi_A + \psi_{AB})$. A non-detection at slit $A$ means that the path must have gone through slit $B$ only ($\psi_B$). Thus the probability density when there is a type I which-path detector at slit $A$ is 
\begin{equation}
P_{D_A} = |\psi_A + \psi_{AB}|^2 + |\psi_B|^2~;
\label{eq:PDA}
\end{equation}
and similarly when there is a which-way detector at slit B [Fig.~\ref{fig:double_slit}(c): setup 3] the probability density is,
\begin{equation}
P_{D_B} = |\psi_B + \psi_{AB}|^2 + |\psi_A|^2~.
\label{eq:PDB}
\end{equation}

Now if we place a second type I detector at slit $B$, three types of paths are distinctly detected: paths that go through slit $A$ ($\psi_A)$  or $B$ ($\psi_B$) only, and non-classical paths that go through both before hitting the detection screen ($\psi_{AB}$). In our \textit{gedanken} experiment this is represented by placing two distinguishable (red and black) balls, one in each slit [Fig.~\ref{fig:double_slit}(d): setup 4].  Looking into the tray we will see a black ball, a red ball, or both balls, to reveal that the particle had gone through slit \textit{A}, \textit{B}, or both, respectively. The probability density when there are type I detectors in both slits is therefore, 
\begin{equation}
P_{D_AD_B} = |\psi_A|^2 + |\psi_B|^2 + |\psi_{AB}|^2 ~.
\label{eq:PDADB}
\end{equation}

When there is a type II which-path detector [Fig.~\ref{fig:double_slit}(e): setup 5] one is not able to distinguish paths that contribute to $\psi_A$ from $\psi_B$. However we can distinguish paths that went through one slit from paths that went through both slits before hitting the detecting screen. The probability density with a type II detector is
\begin{equation}
	P_{D_{AB}} = |\psi_A + \psi_B|^2 + |\psi_{AB}|^2 ~.
\label{eq:PDAB}
\end{equation}

We would like now to quantify the difference in the probability distribution of the single (setup 2) and double (setup 4) which-way detector double-slit experiments. Let the source be at position $r_S=(-S,0)$, the screen detector at $r_D=(D,y_D)$, and slit centers are at $(0,\pm d/2)$ (Fig.~\ref{fig:double_slit}). The slits have $w$ width. The setup is symmetric in the $z$-direction, which means we can ignore this component as it only introduces an irrelevant constant, effectively becoming a 2-dimensional problem~\cite{sawant14}. We assume a monochromatic point source and consider a duration much larger than the time of flight, so that the probability distribution at the screen detector can be quantified with the time-independent 2-point propagator, which is attained by summing over all possible paths between $\mathbf{r}_1$ and $\mathbf{r}_2$,
\begin{equation}
	K(\mathbf{r}_1,\mathbf{r}_2) = \int\mathcal{D}[\mathbf{x}(s)]\exp(ik\int ds) ~,
\label{eq:K1}
\end{equation}
where $\mathcal{D}[\mathbf{x}(s)]$ is the usual path integral measure of paths $\mathbf{x}(s)$ with contour length $s$.  However the problem of summing over all possible paths with the boundary conditions imposed by the slit plane is unwieldy and has yet to be exactly solved. Nevertheless, Sawant \textit{et al.}~\cite{sawant14} argues that a good approximation can be achieved by considering two types of paths: paths with straight trajectory segments from source to slit, then to detecting screen; and paths composing of straight  trajectory segment from source to slit, then to another slit, before hitting the detection screen (Fig.~\ref{fig:double_slit}) - this conjecture has subsequently been supported with finite-difference time-domain simulations (FDTD)~\cite{sinha15}. Following the convention of Sawant \textit{et al.} we will call the former classical paths, and the latter non-classical paths (as we have been doing). One notes that consideration of just the classical paths directly leads to Fresnel's theory of diffraction by a slit~\cite{kumar91}. Using the following free propagator for straight paths~\cite{landau75}
\begin{equation}
	K(\mathbf{r}_1,\mathbf{r}_1) = \frac{k}{2\pi i}\frac{e^{ik|\mathbf{r}_1-\mathbf{r}_2|}}{|\mathbf{r}_1-\mathbf{r}_2|} ~,
\label{eq:K2}
\end{equation}
and the identity
\begin{equation}
	K(\mathbf{r}_1,\mathbf{r}_3) = \int d\mathbf{r}_2K(\mathbf{r}_1,\mathbf{r}_2)K(\mathbf{r}_2,\mathbf{r}_3)~,
\label{eq:K3}
\end{equation}
the propagator for classical paths is 
\begin{equation}
	K_P(\mathbf{r}_D,\mathbf{r}_S) = -\Bigl( \frac{k}{2\pi i}\Bigr)^2 \int dy\frac{e^{ik|l_1+l_2|}}{l_1l_2} ~,
\label{eq:KP1}
\end{equation}
where $l_1=(y^2+S^2)^{1/2}$, $l_2=[(y_D-y)^2+D^2]^{1/2}$, and the integral runs over the width space of slit $P$. We will work in the Fraunhofer limit where the distance from the slit to the source and screen detector is much larger than the slit spacing, so that $l_1\approx S+y^2/2S$ and $l_2\approx D+(y_D-y)^2/2D$, giving ($\gamma\equiv \exp[ik(S+D)]/SD$)~\cite{sawant14}
\begin{equation}
	K_P(\mathbf{r}_D,\mathbf{r}_S) \approx -\gamma\Bigl( \frac{k}{2\pi i}\Bigr)^2 \int dy e^{ik(\frac{y^2}{2S}+\frac{(y_D-y)^2}{2D})} ~.
\label{eq:KP2}
\end{equation}

The propagator for non-classical paths is 
\begin{equation}
K_{PQ}(\mathbf{r}_D,\mathbf{r}_S) = \Bigl( \frac{k}{2\pi i}\Bigr)^3 \int dy_P dy_Q\frac{e^{ik|l_1+l_2+l_3|}}{l_1l_2l_3} ~,
\label{eq:KPQ1}
\end{equation}
where $l_1=(y_P^2+S^2)^{1/2}$, $l_2=y_Q-y_P$, $l_3=[(y_D-y)^2+D^2]^{1/2}$, and the $y_P(y_Q)$ integral runs over the width space of slit $P(Q)$. In the Fraunhofer limit and under the stationary phase approximation~\cite{sawant14},
\begin{equation}
\begin{split}
	K_{PQ}(\mathbf{r}_D,\mathbf{r}_S) \approx &\gamma i^\frac32\Bigl(\frac{k}{2\pi}\Bigr)^\frac52 \int dy_P dy_Q|y_Q-y_P|^{-\frac12}\\
	&\quad\times e^{ik(\frac{y^2}{2S}+|y_Q-y_P|+\frac{(y_D-y)^2}{2D})} ~.
\end{split}
\label{eq:KPQ2}
\end{equation}
The Fraunhofer limit and stationary phase approximation introduce uncertainty on the order of $K\times10^{-4}$~\cite{sawant14}. 

The contributions from non-classical paths become more pronounced as the operating wavelength increases relative to slit spacing. The reason for this is that longer wavelengths mean more overlap between the single-slit wave functions, so that non-classical paths which enter both slits are more likely. For this reason, the recent experiment which measured a non-zero Sorkin parameter worked in the microwave regime. Here as case-study we will consider the optical regime with the following parameters: photon source of $\lambda = 810$~nm wavelength, slit width $w = 500~$nm, inter-slit spacing of $d = 2000~$nm, and source and detector distances $S = D = 1$ mm.

\begin{figure*}
	\centering
	\includegraphics[width=1.75\columnwidth]{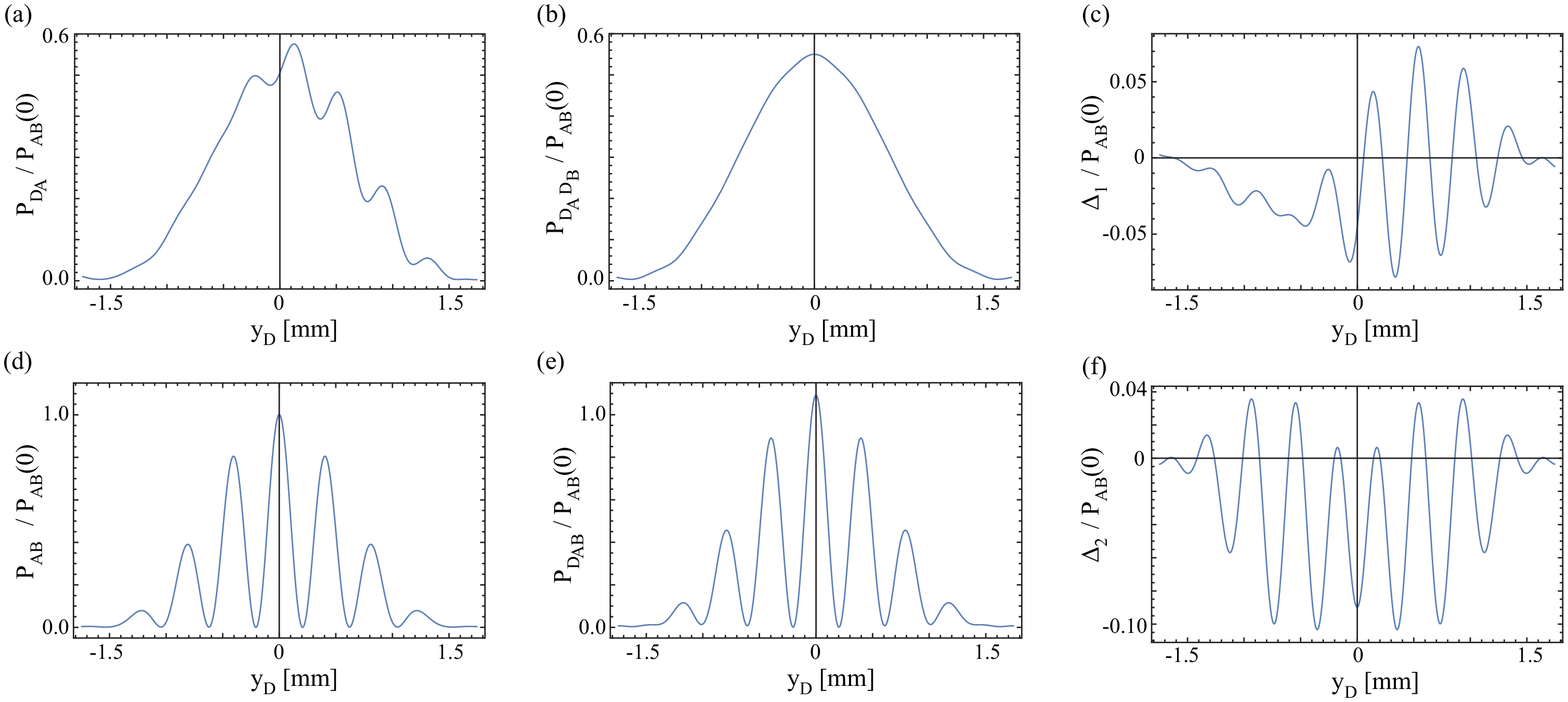}
	\caption{(color online) (a) plots the normalised intensity of the single type I which-way detector experiment (setup 2). (b) plots the normalised intensity of the double type I which-way detector experiment (setup 4). (c) plots the difference in the interference pattern produced by the single and double type I which-way detector experiments, $\Delta_1\equiv P_{D_A}-P_{D_A D_B}$. (d) plots the normalised intensity of the double-slit experiment without which-way detectors (setup 1). (e) plots the normalised intensity of the type II which-way detector experiment (setup 5). (e) plots the difference in the interference pattern between setup 1 and setup 5, $\Delta_2\equiv P_{AB}-P_{D_{AB}}$. Parameters: photon source of $\lambda = 810$~nm wavelength, slit width $w = 500~$nm, inter-slit spacing of $d = 2000~$nm, and source and detector distances $S = D = 1$ mm. All plots are normalised to the maximum central intensity of the double-slit experiment, $P_{AB}(0)$.}
	\label{fig:plot}
\end{figure*}

For a point source at $\mathbf{r}_S$, $\psi (\mathbf{r}_D)=K (\mathbf{r}_D,\mathbf{r}_S$), where $K (\mathbf{r}_D,\mathbf{r}_S)$ is the corresponding propagator from source to screen. Using Eq.~\ref{eq:KP2} and Eq.~\ref{eq:KPQ2}, Fig.~\ref{fig:plot}(a) plots the intensity of the single which-way detector experiment (setup 2), $P_{D_A}$; all plots are normalised to the maximum central intensity of the double-slit experiment, $P_{AB}(0)$. ($P_{D_B}$ is not shown as it is the same as Fig.~\ref{fig:plot}(a) reflected about the $y$-axis.) Fig.~\ref{fig:plot}(b) plots the intensity of the double which-way detector experiment (setup 4), $P_{D_AD_B}$. Fig.~\ref{fig:plot}(c) shows the difference in the interference pattern produced by the single and double which-way detector setups, $\Delta_1\equiv P_{D_A}-P_{D_A D_B}$.

The non-zero value of $\Delta_1=\psi_A\psi_{AB}^*+\psi_A^*\psi_{AB}$ is the result of the interference between the single slit $A$ and the non-classical path wave functions. $\Delta_1/P_{AB}(0)$ is on the order of $10^{-2}$, which is much larger than the uncertainty introduced by the Fraunhoefer limit and stationary phase approximation. In the optical regime, the Sinha \textit{et al.} experiment achieved intensity accuracies of order $10^{-2}$ normalised to the expected two-path interference, and subsequent experiments with multi-path interferometers~\cite{sollner12,kauten16} have  claimed accuracies of at least an order of magnitude better. If the same sort of accuracy can be achieved in double slit experiments with which-way detectors, the effects of the second which-way detector on the interference pattern due to non-classical paths, should be detectable. 

For completeness we have also plotted the intensities of $P_{AB}$ and $P_{D_{AB}}$, and their difference, $\Delta_2$, in Fig.~\ref{fig:plot}(d)-(f). The non-zero value of $\Delta_2=(\psi_A+\psi_B)\psi_{AB}^*+(\psi_A+\psi_B)^*\psi_{AB}$ is the result of the interference between the classical and non-classical path wave functions of the double slits.

\subsection{Born rule violation}

The Sorkin parameter for the triple-slit experiment is defined as
\begin{equation}
	\mathcal{I}_{ABC} \equiv \mathcal{P}_{ABC}-\mathcal{P}_{AB}-\mathcal{P}_{AC}-\mathcal{P}_{BC}+\mathcal{P}_{A}+\mathcal{P}_{B}+\mathcal{P}_{C}~,
\label{eq:IABC}
\end{equation}
where $\mathcal{P}_{ABC}$ is the probability of detection when all 3 slits ($A,B,C$) are open,  $\mathcal{P}_{AB}$ is the probability of detection when 2 slits ($A,B$) are open, and so on. If one assumes that the probabilities are simply given by the linear superposition of the individual wave functions of the constituent single-slit setups ($\mathcal{P}_{ABC}=|\psi_A+\psi_B+\psi_C |^2$, $\mathcal{P}_{AB}=|\psi_A+\psi_B |^2$, and so on), then by rewriting probabilities in Eq.~(\ref{eq:IABC}) in terms of wave functions, it can be shown that $\mathcal{I}_{ABC}=0$ if the Born rule is correct. The proposed advantage of using the Sorkin parameter to experimentally test the Born rule is that one does not need to know the theoretical values of these probabilites, one need only measure them.

If one accounts for non-classical paths, the probability of detection must be corrected to $\mathcal{P}_{ABC}=|\psi_A+\psi_B+\psi_C+\psi_{ABC} |^2$, where $\psi_{ABC}$ is the wave function made up of non-classical paths when slits $A,B,C$ are open, which are not accounted for by single-slit wave functions $\psi_A,\psi_B, \psi_C$. Similar corrections are required for two slits, e.g.  $\mathcal{P}_{AB}=|\psi_A+\psi_B+\psi_{AB} |^2$. The inclusion of these corrections mean that $I_{ABC}\neq 0$.  This was first noted by De Raedt \textit{et al.}~\cite{raedt12}. As the interference pattern of the triple-slit experiment can be described classically, Raedt \textit{et al.} solved Maxwell's equation with FDTD simulations to show the linear single-slit wave function superposition assumption underlying the Sorkin parameter was not correct - therefore $I_{ABC}\neq 0$ does not immediately signal Born rule violation.

In principle one may theoretically calculate this non-zero value of  $I_{ABC}$ with the higher-order corrections. In practice however, there are an infinite number of non-classical paths and an exact calculation is currently impossible (work in this field has produced approximate closed-form solutions~\cite{sinha15}). Having to theoretically calculate probabilities negates the main benefit of using the Sorkin parameter, which is to test the Born rule without the need for such calculations.

Furthermore there are regimes where non-classical paths can have significant contributions, as shown in Fig.~\ref{fig:plot}, and therefore produce relatively large values of the Sorkin parameter even though the Born rule has not been violated. Here we introduce an alternative parameter which will produce exactly zero in all regimes when the Born rule is not violated.  

The Sorkin parameter can be generalised to systems with 3 and more slits, but not 2 slits. The reason for this is that $\mathcal{I}_{AB}\equiv \mathcal{P}_{AB}-\mathcal{P}_A-\mathcal{P}_B\neq 0$ even if one ignores the non-classical paths. This is why the triple-slit experiment is the simplest setup to  test the Born rule with the Sorkin parameter. However if one adds which-way detectors, double-slit experiments can be used to exactly test the Born rule. In particular we introduce the following parameter as an exact test of the Born rule:
\begin{equation}
	I_{AB} \equiv P_{AB}-P_{D_A}-P_{D_B}-P_{D_{AB}}+2P_{D_AD_B}~.
\label{eq:IAB}
\end{equation}

Substitution of Eq.~(\ref{eq:PAB})-(\ref{eq:PDAB}) into Eq.~(\ref{eq:IAB}) shows that $I_{AB}=0$. Like the Sorkin parameter, $I_{AB}$ subtracts from $P_{AB}$ all possible combinations of pair-wise interactions terms yielding $I_{AB}=0$ if the Born rule holds; if the probability of detection is anything other than the absolute square of the wave function, then $I_{AB}\neq 0$ in general. Different from the Sorkin parameter however, $I_{AB}$ exactly accounts for the higher-order corrections to all orders. This allows an exact direct test of the Born rule, limited only to experimental uncertainty. 

Specifically, the experiment to test the Born rule will involve repeating the double-slit experiments 5 times, each time the only thing to be changed is the which-path detector configuration, to get values for $P_{AB},~P_{D_A},~P_{D_B},~P_{D_{AB}}$, and $P_{D_AD_B}$. One may then measure I$_{AB}$ using Eq.~(\ref{eq:IAB}): any value other than $I_{AB}=0$ signals Born rule violation.

\section{Imperfect Detectors}
\label{sec:imperfect}

In practice detectors are imperfect, whether by design or because of technical limitations. Which-way detectors with controllable efficiency have been used in experiments to reveal the quantum-classical boundary~\cite{bertet01,sonnentag07,frabboni10}. Unlike perfect detectors, imperfect detectors introduce detection efficiency as additional parameters. Here we will consider the case of detectors with the same efficiency. For a case-study we will use the parameters set in Sec. II.

To study the double-slit experiment with imperfect detectors we write down the basis independent representation of the wave functions in the previous section as $|\psi_A\rangle$, $|\psi_B\rangle$, and $|\psi_{AB}\rangle$ (we will project to the position basis in the end). When there are no detectors the state of the particle at the detection screen is
\begin{equation}
	|\psi\rangle=|\psi_A\rangle + |\psi_B\rangle + |\psi_{AB}\rangle~.
\label{eq:psi}
\end{equation}

In the presence of detectors, the particle becomes entangled with the detector as it pass through the slits. We  denote the normalised triggered state of a type I detector at slit $A$ (setup 2) as $|D_A\rangle$ and the normalised untriggered state as $|0\rangle$. In terms of the \textit{gedanken} experiment, these states of the detector is represented by 1 and 0 balls in the tray respectively. The state of the system after the particle passes through the slit plane is
\begin{equation}
	|\phi_{D_A}\rangle=(|\psi_A\rangle + |\psi_{AB}\rangle)|1_A\rangle + |\psi_B\rangle|0\rangle~.
\label{eq:psiDA}
\end{equation}
We are only interested in the probability of detection of the particle at the detection screen, so we trace out the detector states from the density matrix obtained from the pure state of Eq.~(\ref{eq:psiDA}). We project this reduced density matrix onto the position basis to get the probability distribution,
\begin{equation}
	P_{D_A}' = |\psi_A+\psi_{AB}|^2+|\psi_B|^2+2\text{Re}[(\psi_A+\psi_{AB})^*\psi_B]\langle0|D_A\rangle~.
\label{eq:PDAp1}
\end{equation}
$\langle0|D_A\rangle$ is the amount of overlap between the triggered and untriggered detector states, which determines the level of interference in the probability distribution. When these states are orthogonal, one retrieves the prefect detector probability distribution of Eq.~(\ref{eq:PDA}). An operative understanding of the overlap term is made clear by setting $\langle0|D_A\rangle=1-n$, where $0\le n\le 1$, and rewriting Eq.~(\ref{eq:PDAp1}) as
\begin{equation}
	P_{D_A}' = nP_{D_A}+(1-n)P_{AB}~.
\label{eq:PDAp2}
\end{equation}
From Eq.~(\ref{eq:PDAp2}) we can interpret $n$ as the efficiency of the detector at slit $A$: the detector will detect an event with efficiency $n$ and when it does the probability distribution is $P_{D_A}$, and the detector will miss an event with efficiency $1-n$ and when it does the probability distribution is $P_{AB}$. Similarly for a detector at slit $B$ only, one gets

\begin{equation}
P_{D_B}' = nP_{D_B}+(1-n)P_{AB}~.
\label{eq:PDBp2}
\end{equation}

For a type I detector at each of the slits (setup 3), we denote the normalised states of the detector system as $|D_A\rangle$, $|D_B\rangle$, $|D_AD_B\rangle$ (representing the black, red, and black and red ball states in the \textit{gedanken} experiment). The state of the system after the slit plane is
\begin{equation}
|\phi_{D_AD_B}\rangle=|\psi_A\rangle|D_{A}\rangle + |\psi_B\rangle|D_{B}\rangle + |\psi_{AB}\rangle|D_{A}D_{B}\rangle~.
\label{eq:psiDADB}
\end{equation}
The corresponding probability distribution is
\begin{equation}
\begin{split}
	P_{D_AD_B}' &= |\psi_A|^2+|\psi_B|^2+|\psi_{AB}|^2\\
		&\quad+2\text{Re}[\psi_A^*\psi_B\langle D_B|D_A\rangle+\psi_A^*\psi_{AB}\langle D_AD_B|D_A\rangle\\
		&\quad+\psi_B^*\psi_{AB}\langle D_AD_B|D_B\rangle]~.
\end{split}
\label{eq:PDADBp1}
\end{equation}
Eq.~(\ref{eq:PDADBp1}) is a general representation of the probability distribution. Let us consider the case where $\langle D_AD_B|D_A\rangle=\langle D_AD_B|D_B\rangle=1-n$ and $\langle D_B|D_A\rangle=(1-n)^2$. This yields
\begin{equation}
	P_{D_AD_B}' = n^2P_{D_AD_B}+n(1-n)(P_{D_A}+P_{D_{B}})+(1-n)^2P_{AB}~.
\label{eq:PDADBp2}
\end{equation}
Our choice of detector-state overlaps can thus be interpreted as the result of two $n$-efficient type-I which-way detectors at slit $A$ and $B$, with  $n^2$ probability that both detectors can detect an event, $n(1-n)$ probability that one detector can detect an event and the other does not, $(1-n)^2$ probability that both detectors missed an event. 

A comparison of Eq.~(\ref{eq:PDAp1}) and Eq.~(\ref{eq:PDADBp1}) shows that even if non-classical paths are neglected (i.e. $\psi_{AB}=0$), the presence of a second which-way detector can have an effect on the probability distribution when the detectors are imperfect; formally the difference results from the fact that $\langle 0|D_A\rangle\neq\langle D_B|D_A\rangle$ in general.

Note that unlike Eq.~(\ref{eq:PDAp2}) and (\ref{eq:PDBp2}), Eq.~(\ref{eq:PDADBp2}) is implementation specific, dependent on the form of the overlap of detector states.

For the type-II detector (setup 4), we denote the normalised states of the detector system as $|D_1\rangle$ and $|D_2\rangle$ (representing the one and two indistinguishable-ball states in the \textit{gedanken} experiment). The state of the system after the slit plane is
\begin{equation}
|\phi_{D_{AB}}\rangle=(|\psi_A\rangle + |\psi_B\rangle)|D_1\rangle + |\psi_{AB}\rangle|D_2\rangle~.
\label{eq:psiDADB}
\end{equation}
The corresponding probability distribution is
\begin{equation}
\begin{split}
P_{D_{AB}}' &= |\psi_A|^2+|\psi_B|^2+|\psi_{AB}|^2+2\text{Re}(\psi_A^*\psi_B)\\
&\quad+2\text{Re}(\psi_A^*\psi_{AB}+\psi_B^*\psi_{AB})\langle D_2|D_1\rangle~.
\end{split}
\label{eq:PDABp1}
\end{equation}
$\langle D_2|D_1\rangle$ gives the amount of overlap between the one and two indistinguishable-ball states. When these states are orthogonal, we can be  sure whether the particle has passed through one or two slits. Conversely, when the states completely overlap we have no information on whether the particle has passed through one or two slits, which is equivalent to having no detectors. In between these two extremes, the detector states partially overlap and we have an imperfect type-II which-way detector. Setting $\langle D_2|D_1\rangle=1-n$, Eq.~(\ref{eq:PDABp1}) can be rewritten as
\begin{equation}
	P_{D_{AB}}' = nP_{D_{AB}}+(1-n)P_{AB}~.
\label{eq:PDABp2}
\end{equation}

\begin{figure*} 
	\centering
	\includegraphics[width=2\columnwidth]{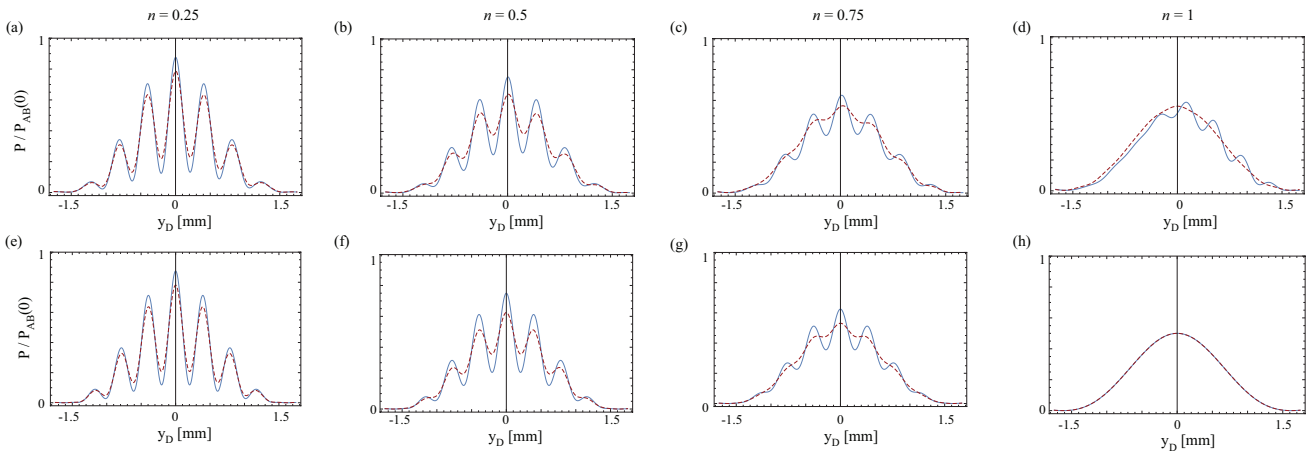}
	\caption{(color online) Comparative plots of single (solid blue line) and double (dashed red line) which-way detectors. (a)-(d) plots $P_{D_A}'$ and $P_{D_AD_B}'$ with higher-order contributions for detector efficiencies $n=0.25,0.5,0.75,1$. (e)-(f) plots $P_{D_A}'$ and $P_{D_AD_B}'$ without higher-order contributions. As the efficiency increases there is a transition from classical wave-like interference to corpuscular quantum behaviour. Parameter values follow Fig.~2.}
	\label{fig:inefficient}
\end{figure*}

Fig.~\ref{fig:inefficient}(a)-(d) plots $P_{D_A}'$ (solid line) and $P_{D_AD_B}'$ (dashed line) for $n=0.25,0.5,0.75,1$ under the parameters set in Fig.~2. As the efficiency increases there is a transition from classical wave-like interference, to corpuscular quantum behaviour due to which-way detector observation. The presence of a second which-way detector increases the efficiency of detection, thereby seeing an earlier transition to corpuscular behaviour.

For comparison we also plot $P_{D_A}'$ (solid line) and $P_{D_AD_B}'$ (dashed line) when the higher-order contributions are ignored (i.e. $\psi_{AB}=0$) in Fig.~\ref{fig:inefficient}(e)-(h). As detector error is eliminated, detection efficiency is eliminated as a parameter which can distinguish the presence of a second detector. If higher-order contributions are neglected, then there is no difference between having one or two perfect detectors, as shown in Fig.~\ref{fig:inefficient}(h). In contrast, if one accounts for higher-order contributions, even as detector error is eliminated, the presence of a second detector cannot be made redundant  [Fig.~\ref{fig:inefficient}(d)], as discussed in the Sec. II.

\begin{figure}
	\centering
	\includegraphics[width=1\columnwidth]{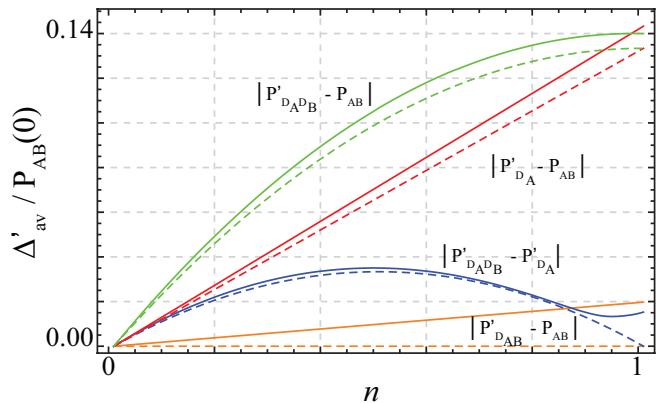}
	\caption{(color online) Average absolute difference in probability distributions as function of which-way detector efficiency. The solid (dashed) line plots the average difference in probability distribution $\Delta_\text{av}'$, with (without) higher-order contributions, between $y_1=-1.75~\text{mm}$ and $y_2=1.75~\text{mm}$. Other parameter values follow Fig.~2.}
	\label{fig:Delta}
\end{figure}

As which-way detectors are not perfect, in practice it is more likely that $P_{D_{A}}',P_{D_{B}}',P_{D_{A}D_B}',P_{D_{AB}}'$ will be the measured quantities. Therefore, by simultaneously solving Eq.~(\ref{eq:PDAp2}), (\ref{eq:PDBp2}), (\ref{eq:PDADBp2}), and (\ref{eq:PDABp2}), we write here the perfect ($n=1$) probability distributions as functions of the imperfect probability distributions ($0<n<1$):
\begin{equation}
P_{D_A} = \frac{P_{D_A}'-(1-n)P_{AB}}{n}~,
\label{eq:PDApp}
\end{equation}
\begin{equation}
P_{D_B} = \frac{P_{D_B}'-(1-n)P_{AB}}{n}~,
\label{eq:PDBpp}
\end{equation}
\begin{equation}
P_{D_AD_B} = \frac{P_{D_AD_B}'+(1-n)^2P_{AB}-(1-n)(P_{D_A}'+P_{D_B}')}{n^2}~,
\label{eq:PDADBpp}
\end{equation}
\begin{equation}
P_{D_{AB}} = \frac{P_{D_{AB}}'-(1-n)P_{AB}}{n}~,
\label{eq:PDABpp}
\end{equation}
Substitution of Eq.~(\ref{eq:PDApp})-(\ref{eq:PDABpp}) into Eq.~(\ref{eq:IAB}) gives the parameter to test for Born rule violation in terms of the measured probability distributions with inefficient which-way detectors. Using this substitution one may then test the Born rule with inefficient detectors. To retrieve the perfect probability distributions from the imperfect ones however, requires resolutions that can distinguish the different imperfect probability distributions. 

Fig.~\ref{fig:Delta} plots the average absolute difference in the probability distribution as a function of detector efficiency for the various probability functions of Eq.~(\ref{eq:PDAp2}), (\ref{eq:PDBp2}), (\ref{eq:PDADBp2}), and (\ref{eq:PDABp2}), with higher-order contributions (solid line) and without higher-order contributions (dashed line),
\begin{equation}
\Delta_\text{av}'=\frac{1}{y_2-y_1}\int_{y_1}^{y_2}|P_P'-P_Q'|~dy~.
\label{eq:Deltap}
\end{equation} 
Overall the lower the efficiency of the detectors, the harder it is to distinguish the different imperfect probability distributions, with the difference between  $P_{D_{A}}'$ and $P_{D_{A}D_B}'$ (solid blue line), and $P_{D_{AB}}$ and $P_{AB}'$ (solid orange line) being the most difficult to observe. [Note that Fig.~(\ref{fig:inefficient}) corresponds to the  blue lines in Fig.~(\ref{fig:Delta}).] Under the experimental parameters used in Fig.~(\ref{fig:plot}), if one were to achieve an accuracy of $10^{-2}$ in the measurement of the probability distributions (as achieved by the Sinha \textit{et al.} experiment~\cite{sinha10}), then Fig.~\ref{fig:Delta} shows that the required which-way detector efficiency to distinguish the different imperfect probability distributions would need to be greater than 50\%.

\section{Conclusion}

We have shown that the inclusion of higher-order or non-classical paths will lead to different interference patterns for which-way double-slit experiments with one and two which-way detectors. These differences should be measurable in regimes where the operating wavelength is commensurate to or larger than slit spacing. Previously, direct tests of the Born rule have been triple-slit experiments measuring the Sorkin parameter. The Sorkin parameter however is only an approximate test of the Born rule, and can only be applied in regimes where the operating wavelength is much smaller than slit spacing. By explicitly accounting for higher-order correction, we have given an alternative parameter which is an exact test of the Born rule for all wavelengths and slit spacing.  This should open up a new suite of experiments based on which-path double-slit experiments, to test the Born rule to accuracies limited only by experimental uncertainties and not theoretical ones.   

\section{Acknowledgements}
The author would like to thank M. Lajk\'{o}, A. Celi, C.-H. Su,  and S. Quach for discussions and checking the manuscript. This work was financially supported by the European Commission's Marie Curie Actions for Co-funding of Regional, National and International Programmes (COFUND), the Spanish Ministry of Economy and Competitiveness (MINECO), Fundacio Privada Cellex, Spanish Ministerio de Econom\'{i}a y Competitividad grants (Severo Ochoa SEV-2015-0522 and FOQUS FIS2013-46768-P), and Generalitat de Catalunya (2014 SGR 874).

\bibliography{born_prl}

\end{document}